# A New Methodology of Spatial Crosscorrelation Analysis

Yanguang Chen

Department of Geography, College of Urban and Environmental Sciences, Peking University, 100871, Beijing, China. Email: chenyg@pku.edu.cn

**Abstract**: The idea of spatial crosscorrelation was conceived of long ago. However, unlike the related spatial autocorrelation, the theory and method of spatial crosscorrelation analysis have remained undeveloped. This paper presents a set of models and working methods for spatial crosscorrelation analysis. By analogy with Moran's index newly expressed in a spatial quadratic form and by means of mathematical reasoning, I derive a theoretical framework for geographical crosscorrelation analysis. First, two sets of spatial crosscorrelation coefficients are defined, including a global spatial crosscorrelation coefficient and a set of local spatial crosscorrelation coefficients. Second, a pair of scatterplots of spatial crosscorrelation is proposed, and different scatterplots show different relationships between correlated variables. Based on the spatial crosscorrelation coefficient, Pearson's correlation coefficient can be decomposed into two parts: direct correlation (partial crosscorrelation) and indirect correlation (spatial crosscorrelation). As an example, the analytical process is applied to the relationships between China's urbanization and economic development. Spatial crosscorrelation and spatial autocorrelation can complement one another, and the spatial crosscorrelation scatterplots can be used to reveal the causality inside a self-organized system. The spatial crosscorrelation models will play a useful role in future geographical spatial analysis.

# 1 Introduction

In geographical research, spatial correlation falls into two types: autocorrelation and crosscorrelation. The former reflects intrasample correlation, that is, a relation between one measure and itself, while the latter reflects intersample correlation, namely, a relationship between one measure and another measure. Spatial autocorrelation is defined by one size measurement (e.g. city population) and one spatial contiguity measurement (e.g., Euclidean distance), while spatial crosscorrelation can be defined by two size measurements (e.g., city population and urban area) and one spatial contiguity measurement. Based on the statistical measurements of Moran's index and Geary's coefficient (Moran, 1950; Geary, 1954), a relatively mature theory has been developed for spatial autocorrelation (Anselin, 1995; Anselin, 1996; Bivand *et al*, 2009; Chen, 2012; Chen, 2013; Cliff and Ord, 1969; Cliff and Ord, 1973; Cliff and Ord, 1981; Fischer and Wang, 2011; Getis, 2009; Getis and Ord, 1992; Griffith, 2003; Haggett *et al*, 1977; Haining, 2009; Jiang and Yao, 2010; Li *et al*, 2007; Odland, 1988; Sokal and Oden, 1978; Sokal and Thomson, 1987; Tiefelsdorf , 2002; Wall, 2004; Wang *et al*, 2012; Weeks *et al*, 2004). Spatial autocorrelation analysis has been widely applied to various correlational analyses of natural and human phenomena in many fields (Beck and Sieber, 2010; Benedetti-Cecchi *et al*, 2010; Bizzarro *et al*, 2014; Bonnot *et al*, 2010; Braun *et al*, 2012; Chu *et al*, 2013; Deblauwe *et al*, 2012; Dore *et al*, 2014; Impoinvil *et al*, 2011; Kumar *et al*; 2012; Lichstein *et al*, 2002; Mateo-Tomás and Olea, 2010; Mattsson *et al*, 2013; Stark *et al*, 2012; Wang, 2006; Wang *et al*, 2011), and in particular it has been integrated into the spatial analytical technology of geographical information systems (GIS) (Longley *et al*, 2011; Smith *et al*, 2009). In contrast, the theory and methodology of spatial crosscorrelation has not yet been well constructed for geographical analysis, despite the concept of "spatial crosscorrelation" emerging in literature (Chen, 2009; Duffy and Hughes-Clarke, 2005; Kleiber and Genton, 2013; Lamb *et al*, 2014; Longley and Batty, 1996; Loth and Baker, 2013; Rack *et al*, 2008).

For a number of geographical elements within a regional system, the relationship between two measurements used to be characterized with Pearson's correlation coefficient, which indicates the simplest crosscorrelation. However, Pearson's correlation coefficient shows nothing about interactions based on spatial distances. In this paper, I present a new graceful theoretical framework for spatial correlation analysis. The novelty of this framework rests with three aspects. First, it is found by analogy with a new expression of Moran's index (Chen, 2013). Therefore, the definition of spatial crosscorrelation coefficient is easy to understand, and the relationship between spatial autocorrelation and spatial crosscorrelation is clear. Second, it is formulated in the simplest form of vectors and matrices, so it is easy to calculate the correlation coefficient using MS Excel or Matlab. Third, the procedure of calculations and analysis is well developed. The methodological framework contains global indices, local indices, and crosscorrelation scatterplots.

The rest of the article is arranged as follows. In Section 2 (*Results*), the global and local indices of spatial crosscorrelation are defined, and a pair of spatial crosscorrelation scatterplots is presented by analogy with Moran's scatterplots. In Section 3 (*Discussion*), Pearson's correlation coefficient is decomposed into a direct correlation coefficient and an indirect correlation coefficient based on the idea from spatial crosscorrelation given in Section 2, and a comparison is drawn between the spatial crosscorrelation coefficient and Moran's index. In Section 4 (*Materials and Methods*), as a case study, the analytical process of spatial crosscorrelation is applied to the system of China's cities and regions. Finally, the paper concludes with a summary of the main points of this work.

## 2 Models

### 2.1 Global and local measurements of spatial crosscorrelation

The theoretical framework consists of a set of models, and the related mathematical reasoning will be shown first. Suppose there are $n$ elements (e.g., cities) in a system (e.g., a network of cities) which can be measured by two variables (e.g., city population and urban area), $X$ and $Y$. A pair of vectors can be defined as below:

$$X = \begin{bmatrix} x_1 & x_2 & \cdots & x_n \end{bmatrix}^\mathrm{T}, \quad Y = \begin{bmatrix} y_1 & y_2 & \cdots & y_n \end{bmatrix}^\mathrm{T}, \tag{1}$$

where $x_i$ and $y_i$ are two size measurements of the $i$th element ($i=1,2,\ldots,n$), and the symbol "T" denotes transpose, a process of interchanging the rows and columns of a given matrix. The means of $x_i$ and $y_i$ are given as follows

$$\mu_x = \frac{1}{n}\sum_{i=1}^{n} x_i, \mu_y = \frac{1}{n}\sum_{i=1}^{n} y_i. \tag{2}$$

The centralized variable can be calculated by

$$X_c = X - \mu_x, Y_c = Y - \mu_y \tag{3}$$

where $\mu_x$ and $\mu_y$ represent the average values of the variables $x_i$ and $y_i$. The standard deviations of the two variables are as follows

$$\sigma_x^2 = \frac{1}{n}\sum_{i=1}^{n}(x_i - \mu_x)^2 = \frac{1}{n}X_c^T X_c, \sigma_y^2 = \frac{1}{n}\sum_{i=1}^{n}(y_i - \mu_y)^2 = \frac{1}{n}Y_c^T Y_c, \tag{4}$$

where $\sigma_x^2$ and $\sigma_y^2$ are the population variances (PVs) of $x_i$ and $y_i$, respectively. The results of a scaling transform of the centralized variables form a pair of standardized vectors such as

$$x = \frac{X - \mu_x}{\sigma_x} = \frac{X_c}{\sigma_x}, y = \frac{Y - \mu_y}{\sigma_y} = \frac{Y_c}{\sigma_y}. \tag{5}$$

which are termed *standard scores* in statistics. It can be shown that the norm of $x$ and $y$, i.e., the lengths of the vectors, $\|x\|$ and $\|y\|$, exactly equals the dimensions of the system, i.e., the number of elements in the system, $n$. Thus we have

$$\|x\| = x^T x = n, \|y\| = y^T y = n. \tag{6}$$

The models of spatial correlation, including autocorrelation and crosscorrelation, are based on spatial distance or spatial contiguity. Define an *n*-by-*n* unitary spatial weights matrix such as

$$W = [w_{ij}]_{n \times n}. \tag{7}$$

which is actually a unitized spatial weights matrix (USWM). The matrix can be produced by a spatial contiguity matrix (SCM), and it has three properties as below: (1) Symmetry, i.e., $w_{ij}=w_{ji}$; (2) Zero diagonal elements, namely, $|w_{ii}|=0$, which implies that the entries in the diagonal are all 0; (3) Unitization condition, that is

$$\sum_{i=1}^{n}\sum_{j=1}^{n} w_{ij} = 1. \tag{8}$$

Then, by analogy with the improved formula of Moran's index for spatial autocorrelation (Chen, 2013), a new measurement for spatial crosscorrelation analysis can be defined as

$$R_c = x^T W y, \tag{9}$$

where $R_c$ denotes the coefficient of spatial crosscorrelation, which can be termed *spatial crosscorrelation index* (SCI). It is easy to prove that the SCI is a correlation coefficient, and its value falls between -1 and 1. Because of symmetry of the spatial weights matrix, transposing $R_c$ yields another expression

$$R_c = (x^T W y)^T = y^T W^T x = y^T W x, \tag{10}$$

which is equivalent to equation (9). However, as indicated in the following section, from equations (9) and (10), we can derive different models for different uses of spatial analysis.

A set of matrix equations can be constructed, based on the SCI formulae. Equations (9) and (10) multiplied left by $x$ or $y$ on both sides of the equal signs yields

$$M^{(xy)} x = x y^T W x = R_c x, \tag{11}$$

$$M^{(yx)} y = y x^T W y = R_c y, \tag{12}$$

$$M^{(xx)} y = x x^T W y = R_c x, \tag{13}$$

$$M^{(yy)} x = y y^T W x = R_c y. \tag{14}$$

We can demonstrate that $x y^T W x = x x^T W y$, $y x^T W y = y y^T W x$. In these equations, there are two Ideal Spatial Correlation Matrixes (ISCM) for spatial autocorrelation as follows

$$M^{(xx)} = x x^T W, \quad M^{(yy)} = y y^T W; \tag{15}$$

there are two ISCMs for spatial crosscorrelation such as

$$M^{(xy)} = x y^T W, \quad M^{(yx)} = y x^T W. \tag{16}$$

SCI is just the eigenvalue of the ISCM of spatial crosscorrelation. This differs from Moran's index, which is the characteristic value of the ISCM of spatial autocorrelation (Chen, 2013).

An important measurement of spatial autocorrelation is called Local Indicators of Spatial Association (LISA). LISA is also termed local Moran's index (Anselin, 1995). Similarly, a pair of sets of local spatial crosscorrelation coefficients can be defined by

$$R_i^{(xy)} = x_i \sum_{j=1}^{n} w_{ij} y_j, \tag{17}$$

$$R_j^{(yx)} = y_i \sum_{j=1}^{n} w_{ij} x_j, \tag{18}$$

where $R_i$ and $R_j$ refer to the *local spatial crosscorrelation index* (LSCI) of the *i*th element and the *j*th element. Accordingly, $R_c$ denotes the *global spatial crosscorrelation index* (GSCI), which can be termed SCI for short. As $w_{ij}=w_{ji}$, for arbitrary *n*, equations (17) and (18) can be expressed with matrix equations such as

$$\begin{bmatrix} x_1 \\ x_2 \\ \vdots \\ x_n \end{bmatrix} \begin{bmatrix} y_1 & y_2 & \cdots & y_n \end{bmatrix} \begin{bmatrix} w_{11} & w_{12} & \cdots & w_{1n} \\ w_{21} & w_{22} & \cdots & w_{2n} \\ \vdots & \vdots & \ddots & \vdots \\ w_{n1} & w_{n2} & \cdots & w_{nn} \end{bmatrix} = \begin{bmatrix} x_1 \sum_{j=1}^n w_{1j} y_j & x_1 \sum_{j=1}^n w_{2j} y_j & \cdots & x_1 \sum_{j=1}^n w_{nj} y_j \\ x_2 \sum_{j=1}^n w_{1j} y_j & x_2 \sum_{j=1}^n w_{2j} y_j & \cdots & x_2 \sum_{j=1}^n w_{nj} y_j \\ \vdots & \vdots & \ddots & \vdots \\ x_n \sum_{j=1}^n w_{1j} y_j & x_n \sum_{j=1}^n w_{2j} y_j & \cdots & x_n \sum_{j=1}^n w_{nj} y_j \end{bmatrix}, \quad (19)$$

$$\begin{bmatrix} y_1 \\ y_2 \\ \vdots \\ y_n \end{bmatrix} \begin{bmatrix} x_1 & x_2 & \cdots & x_n \end{bmatrix} \begin{bmatrix} w_{11} & w_{12} & \cdots & w_{1n} \\ w_{21} & w_{22} & \cdots & w_{2n} \\ \vdots & \vdots & \ddots & \vdots \\ w_{n1} & w_{n2} & \cdots & w_{nn} \end{bmatrix} = \begin{bmatrix} y_1 \sum_{j=1}^n w_{1j} x_j & y_1 \sum_{j=1}^n w_{2j} x_j & \cdots & y_1 \sum_{j=1}^n w_{nj} x_j \\ y_2 \sum_{j=1}^n w_{1j} x_j & y_2 \sum_{j=1}^n w_{2j} x_j & \cdots & y_2 \sum_{j=1}^n w_{nj} x_j \\ \vdots & \vdots & \ddots & \vdots \\ y_n \sum_{j=1}^n w_{1j} x_j & y_n \sum_{j=1}^n w_{2j} x_j & \cdots & y_n \sum_{j=1}^n w_{nj} x_j \end{bmatrix}. \quad (20)$$

Comparing equations (19) and (20) with equations (17) and (18) shows that the elements in the diagonals of $M^{(xy)}$ and $M^{(yx)}$ give the LSCI values. The traces of $M^{(xy)}$ or $M^{(yx)}$ are equal to the GSCI value. It is very convenient for us to compute the LSCIs by means of matrix operations.

## 2.2 Practical equations for spatial crosscorrelation

In practice, the spatial crosscorrelation coefficient can be defined in another form. The precondition for equation (9) is as follows

$$nWy = R_c x, \quad (21)$$

which represents a practical relation for SCI. In fact, according to equation (6), equation (21) multiplied left by $x^T$ yields $nx^T W y = x^T R x = nR$, which results in equation (9). Similarly, the precondition equation (10) is as below

$$nWx = R_c y, \quad (22)$$

which multiplied left by $y^T$ yields $ny^T W x = y^T R y = nR$, which yields equation (10). A Real Spatial

Correlation Matrix (RSCM) for spatial crosscorrelation can be defined as

$$M = nW = \|x\|W = \|y\|W = x^{\mathrm{T}}xW = y^{\mathrm{T}}yW. \tag{23}$$

It can be proved that $R_c$ is just the eigenvalue of $M$, and the corresponding eigenvector is $(x+y)$. Actually, equation (21) plus equation (22) yields

$$M(x+y) = nW(x+y) = R_c(x+y). \tag{24}$$

This suggests that $M$ corresponds to $M^{(xy)}$ and $M^{(yx)}$. The relationship between equation (13) and equation (21) gives an error equation

$$(M^{(xx)}-M)y = (xx^{\mathrm{T}}W - nW)y = U, \tag{25}$$

in which $U$ represents an error vector. The relationship between equation (14) and equation (22) gives another error equation

$$(M^{(yy)} - M)x = (yy^{\mathrm{T}}W - nW)x = V, \tag{26}$$

in which $V$ represents another error vector. Empirically, there are always errors between $My=x^{\mathrm{T}}xWy$ and $M^{(xx)}y=xx^{\mathrm{T}}Wy$, also there are errors between $Mx=y^{\mathrm{T}}yWx$ and $M^{(yy)}x=yy^{\mathrm{T}}Wx$. This suggests an approach to testing the "goodness of fit" a spatial crosscorrelation analysis. If the spatial crosscorrelation is very strong, $Mx$ will be a very close to $M^{(yy)}x$, and $My$ will be a very close to $M^{(xx)}y$.

## 2.3 Spatial crosscorrelation scatterplots

Spatial crosscorrelation can be visually displayed with two scatterplots, which are similar to Moran's scatterplot of spatial autocorrelation. However, the crosscorrelation scatterplots come in pairs. In order to create the scatterplots, six variables based on the spatial correlation matrix (SCM) are defined as below:

$$f^{(xy)} = M^{(xy)}x = xy^{\mathrm{T}}Wx, \tag{27}$$

$$f^{(yx)} = M^{(yx)}y = yx^{\mathrm{T}}Wy, \tag{28}$$

$$f^{(xx)} = M^{(xx)}y = xx^{\mathrm{T}}Wy, \tag{29}$$

$$f^{(yy)} = M^{(yy)}x = yy^{\mathrm{T}}Wx. \tag{30}$$

$$f^{(y)} = My = nWy. \tag{31}$$

$$f^{(x)} = Mx = nWx. \tag{32}$$

Using these equations, we can generate a set of scatterplots comprising four graphs with observational data and calculations.

The variables can be matched to make crosscorrelation scatterplots as follows. The relationship between $x$ and $f^{(xy)}$ give the first scatterplot, the relationship between $x$ and $f^{(xx)}$ give the second scatterplot, the relationship between $y$ and $f^{(yx)}$ give the third scatterplot, and the relationship between $y$ and $f^{(yy)}$ give the fourth scatterplot (Table 1). In fact, the first plot is the same as the second one, while the third plot is identical in form to the fourth one. Therefore, we actually need two scatterplots for spatial crosscorrelation analysis in empirical studies.

Table 1 The functional relationships of two pairs of scatterplots defined for spatial crosscorrelation analysis

| Scatterplot | Abscissa ($x$-axis) | Ordinate ($y$-axis) | | Effect |
|---|---|---|---|---|
| | | Scattered points | Trend line | |
| **The first plot** | $x$ | $f^{(y)}=nWy$ | $f^{(xy)}=xy^TWx$ | $x$ acts on $y$ |
| **The second plot** | $x$ | $f^{(y)}=nWy$ | $f^{(xx)}=xx^TWy$ | $x$ acts on $y$ |
| **The third plot** | $y$ | $f^{(x)}=nWx$ | $f^{(yx)}=yx^TWy$ | $y$ reacts on $x$ |
| **The fourth plot** | $y$ | $f^{(x)}=nWx$ | $f^{(yy)}=yy^TWx$ | $y$ reacts on $x$ |

The crosscorrelation scatterplots can be easily yielded by the spreadsheet, Microsoft Excel, or the matrix programming language, Matlab. Taking $x$ or $y$ as an abscissa ($x$-axis) and $f^{(y)}$ or $f^{(x)}$ as an ordinate ($y$-axis), we can create a scatterplot. Then using the relationships between $x$ or $y$ and $f^{(xx)}$ or $f^{(xy)}$ or $f^{(yx)}$ or $f^{(yy)}$, we can produce a trendline. Each scatterplot includes two parts: $n$ scattered points and a straight line. The relationship between $x$ or $y$ and $f^{(y)}$ or $f^{(x)}$ take on scattered points, but the relationship between $x$ or $y$ and $f^{(xx)}$ or $f^{(xy)}$ or $f^{(yx)}$ or $f^{(yy)}$ exhibit a trendline, which is in fact a regression line. In other words, the plot of $f^{(y)}$ or $f^{(x)}$ vs. $x$ or $y$ presents a set of randomly scattered data points, while the plot of $f^{(xx)}$ or $f^{(xy)}$ or $f^{(yx)}$ or $f^{(yy)}$ vs. $x$ or $y$ shows a set of ordered data points, which make a straight line. Superimposing the trendline onto the scattered data points yields a scatter diagram for spatial crosscorrelation analysis.

# 3 Discussion

## 3.1 Geographical meanings of spatial crosscorrelation measurements

The geographical meaning of the spatial crosscorrelation can be illuminated by clarifying the mathematical relationship between Peasron's correlation coefficient and the SCI. Leaving out spatial distances, we can re-express equations (9) and (10) as follows

$$R_0 = x^T W_0 y = y^T W_0 x, \tag{33}$$

where $R_0$ is the *simple correlation coefficient* (SCC), which can be treated as a special case of SCI, and

$$W_0 = \frac{1}{n} E \tag{34}$$

represents a unitized identity matrix, which takes the place of the USWM, and $E$ denotes an identity matrix, which is a square matrix with $n$ numerals, 1, along the diagonal from upper left to lower right and $n(n-1)$ numerals, 0, in all other positions. It is easy to prove that $R_0$ is just a Pearson's correlation coefficient:

$$R_0 = x^T (\frac{1}{n} E) y = y^T (\frac{1}{n} E) x = \frac{1}{n} x^T y = \frac{1}{n} y^T x, \tag{35}$$

which indicates simple crosscorrelation between $x$ and $y$. A partial correlation coefficient can be defined as

$$R_p = R_0 - R_c = x^T W_0 y - x^T W y = y^T W_0 x - y^T W x, \tag{36}$$

where $R_p$ refers to the partial *spatial* crosscorrelation coefficient (PSCC).

Now, the meanings of the spatial correlation coefficients can be explained as follows. The SCI, $R_c$, denotes the indirect correlation between $x$ and $y$ through the spatial distance and other elements in a geographical system; the PSCC, $R_p$, represents the direct crosscorrelation between $x$ and $y$; Pearson's correlation coefficient, $R_0$, is a simple crosscorrelation coefficient reflecting the summation of spatial correlation, including both the direct crosscorrelation and the indirect crosscorrelation. The SCI has two functions. First, it presents the indirect correlation between $x$ and $y$, which is based on spatial distance. Second, by means of indirect spatial crosscorrelation coefficient, we can estimate the direct crosscorrelation coefficient. Thus, the simple spatial correlation, Peason's correlation, can be separated into two parts: a direct correlation without distance effect and an indirect correlation based on a distance decay effect.

**3.2 A comparison between spatial autocorrelation and spatial crosscorrelation**

In spatial analysis, autocorrelation and crosscorrelation represent two different sides of the same coin. In fact, the concept of autocorrelation comes from the simplest crosscorrelation, i.e. one without a time lag. The autocorrelation coefficient defined in the 2-dimensional space proceeds from the autocorrelation function defined in the 1-dimensional time or space (Figure 1). The 2-dimensional crosscorrelation coefficient is constructed by analogy with the 2-dimensional autocorrelation coeffient, i.e., Moran's index, which was re-expressed in a new mathematical form (Chen, 2013). A comparison can be drawn between spatial autocorrelation and spatial crosscorrelation as shown in Table 2. In short, the spatial autocorrelation is the intrasample spatial correlation, while the spatial crosscorrelation is the intersample spatial correlation. The former is based on one size measurement, while the latter is based on two size measurements.

Table 2 The similarities and differences between spatial autocorrelation and spatial crosscorrelation

| Item | Spatial autocorrelation | Spatial crosscorrelation |
| --- | --- | --- |
| **Correlation property** | Intrasample correlation | Intersample correlation |
| **Correlation coefficient** | $I_x=x^T W x$, $I_y=y^T W y$ | $R_{xy}=x^T W y$, $R_{yx}=y^T W x$ |
| **ISWM** | $M^{(xx)}=xx^T W$, $M^{(yy)}=yy^T W$ | $M^{(xy)}=xy^T W$, $M^{(yx)}=yx^T W$ |
| **RSWM** | $M=nW=x^T x W=y^T y W$ | $M=nW=x^T x W=y^T y W$ |
| **Scatterplot** | One plot | Two plots |

The 2-dimension spatial correlation analyses, including spatial autocorrelation and spatial crosscorrelation, are based on spatial weight matrices. A spatial weight matrix comes from a spatial contiguity matrix (SCM). We have at least four approaches to make a SCM (Chen, 2012). For a geographical system with *n* spatial elements, A SCM can be expressed as

$$V = [v_{ij}]_{n \times n} = \begin{bmatrix} v_{11} & v_{12} & \cdots & v_{1n} \\ v_{21} & v_{22} & \cdots & v_{2n} \\ \vdots & \vdots & \ddots & \vdots \\ v_{n1} & v_{n2} & \cdots & v_{nn} \end{bmatrix}, \quad (37)$$

where $V$ denotes the SCM, and $v_{ij}$ is a measure used to compare and judge the degree of contiguity between place $i$ and place $j$ ($i, j=1,2,\ldots,n$). The elements on the diagonal are zeros, otherwise they must be turned into zero (i.e., for $i=j$, $v_{ii}=0$). A USWM can be defined as $w_{ij}=v_{ij}/T$, where $T$ denotes the sum of SCM entries, that is

$$T = \sum_{i=1}^{n}\sum_{j=1}^{n} v_{ij}. \quad (38)$$

Thus, based on the *population standard deviation* (PSD), the SCI formulae, equations (9) and (10), can be developed in a sophisticated form as follows

$$R_c = \frac{X^T(nW)Y}{X^T Y} = \frac{n\sum_{i=1}^{n}\sum_{j=1}^{n} v_{ij}(X_i - \mu_x)(Y_j - \mu_y)}{T\sqrt{\sum_{i=1}^{n}(X_i - \mu_x)^2 \sum_{i=1}^{n}(Y_i - \mu_y)^2}}, \quad (39)$$

$$R_c = \frac{Y^T(nW)X}{Y^T X} = \frac{n\sum_{i=1}^{n}\sum_{j=1}^{n} v_{ij}(Y_i - \mu_y)(X_j - \mu_x)}{T\sqrt{\sum_{i=1}^{n}(Y_i - \mu_y)^2 \sum_{i=1}^{n}(X_j - \mu_x)^2}}, \quad (40)$$

which bear an analogy with the traditional expression of Moran's index. If our spatial analysis is based on a sample rather than a population (universe), the PSD should be replaced by the *sample standard deviation* (SSD). In this case, equations (39) and (40) should be revised as below

$$R_c = \frac{X^T[(n-1)W]Y}{X^T Y} = \frac{(n-1)\sum_{i=1}^{n}\sum_{j=1}^{n} v_{ij}(X_i - \mu_x)(Y_j - \mu_y)}{T\sqrt{\sum_{i=1}^{n}(X_i - \mu_x)^2 \sum_{i=1}^{n}(Y_i - \mu_y)^2}}, \quad (41)$$

$$R_c = \frac{Y^T[(n-1)W]X}{Y^T X} = \frac{(n-1)\sum_{i=1}^{n}\sum_{j=1}^{n} v_{ij}(Y_i - \mu_y)(X_j - \mu_x)}{T\sqrt{\sum_{i=1}^{n}(Y_i - \mu_y)^2 \sum_{i=1}^{n}(X_j - \mu_x)^2}}. \quad 42)$$

For the comparability of the spatial crosscorrelation measurement with Moran's index, an empirical analysis will be made using PSD rather than SSD in this paper.

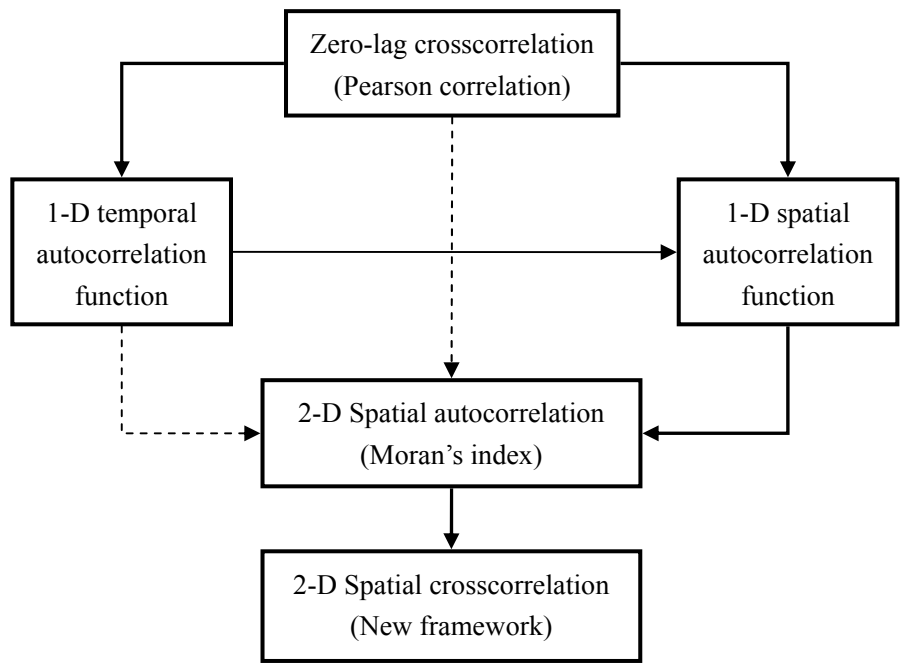

**Figure 1 The paths from simple crosscorrelation to the 2-dimensional spatial crosscorrelation by way of autocorrelation**

(**Note**: In the block diagram, the solid line represents direct relations or paths, while the dashed line denotes the indirect relations or paths. "1-D" refers to "1-dimensional", and "2-D" to "2-dimensional".)

## 4 Materials and Methods

### 4.1 Study area, measurements, and analytical process

The new framework of spatial crosscorrelation can be employed to study the relationship between urbanization and economic development of a country. It proved that there is correlation between population urbanization and regional economic development (Zhou, 1989). However, the relationship between cause and effect is not yet clear. The spatial crosscorrelation analysis can be employed to reveal the causality between urbanization and economic development. As an example, the spatial crosscorrelation models and methods will be applied to Mainland China's regions and cities. The spatial objects are the 31 provinces, autonomous regions, and municipalities directly under the Central Government of China and the capital cities of these regions. The *level of urbanization* is measured by the proportion of urban population to total population in a region, while the level of economic development is measured by the per capita *gross regional product* (GRP). As for the spatial weight matrix, the distances by train between any two capital cities can be used to quantify the spatial contiguity. The statistical data of urbanization levels and per capita GRP (2000-2012) are available from the website of National Bureau of Statistics (NBS) of the People's Republic of China (http://www.stats.gov.cn/tjsj/ndsj/), and the railroad distance matrix can be found in many Chinese road atlases. Because the cities of Haikou and Lhasa were not connected to the network of Chinese cities by railway from 2000 to 2012, only 29 regions and their capital cities are taken into account, and thus the size of each spatial sample is *n*=29 (Table 3).

Table 3 The GRP, level of urbanization, and the LSCI values of 29 Chinese regions (2012)

| Region | City | Original variables | | Standard variables | | LSCI | |
|---|---|---|---|---|---|---|---|
| | | pc GRP ($X$) | Urbanization level ($Y$) | $x$ | $y$ | $xy^\mathrm{T}W$ | $yx^\mathrm{T}W$ |
| Beijing | Beijing | 87475 | 86.20 | 2.1965 | 2.3931 | 0.0384 | 0.0593 |
| Tianjin | Tianjin | 93173 | 81.55 | 2.4875 | 2.0415 | 0.0589 | 0.0485 |
| Hebei | Shijiazhuang | 36584 | 46.80 | -0.4029 | -0.5860 | -0.0061 | -0.0099 |
| Shanxi | Taiyuan | 33628 | 51.26 | -0.5539 | -0.2487 | -0.0021 | -0.0020 |
| Inner Mongolia | Hohehot | 63886 | 57.74 | 0.9916 | 0.2412 | 0.0041 | 0.0009 |
| Liaoning | Shenyang | 56649 | 65.65 | 0.6220 | 0.8393 | 0.0040 | 0.0053 |
| Jilin | Changchun | 43415 | 53.70 | -0.0540 | -0.0642 | -0.0005 | -0.0003 |
| Heilongjiang | Harbin | 35711 | 56.90 | -0.4475 | 0.1777 | -0.0022 | 0.0010 |
| Shanghai | Shanghai | 85373 | 89.30 | 2.0891 | 2.6275 | 0.0099 | 0.0264 |
| Jiangsu | Nanjing | 68347 | 63.00 | 1.2195 | 0.6389 | 0.0134 | 0.0059 |
| Zhejiang | Hangzhou | 63374 | 63.20 | 0.9655 | 0.6541 | 0.0176 | 0.0101 |
| Anhui | Hefei | 28792 | 46.50 | -0.8009 | -0.6086 | -0.0078 | -0.0080 |
| Fujian | Fuzhou | 52763 | 59.60 | 0.4235 | 0.3819 | 0.0003 | 0.0001 |
| Jiangxi | Nanchang | 28800 | 47.51 | -0.8005 | -0.5323 | -0.0018 | -0.0012 |
| Shandong | Jinan | 51768 | 52.43 | 0.3727 | -0.1603 | 0.0045 | -0.0022 |
| Henan | Zhengzhou | 31499 | 42.43 | -0.6626 | -0.9164 | -0.0027 | -0.0042 |
| Hubei | Wuhan | 38572 | 53.50 | -0.3013 | -0.0794 | 0.0009 | 0.0002 |
| Hunan | Changsha | 33480 | 46.65 | -0.5614 | -0.5973 | 0.0001 | 0.0011 |
| Guangdong | Guangzhou | 54095 | 67.40 | 0.4915 | 0.9716 | -0.0011 | -0.0020 |
| Guangxi | Nanning | 27952 | 43.53 | -0.8438 | -0.8332 | 0.0013 | 0.0017 |
| Chongqing | Chongqing | 38914 | 56.98 | -0.2839 | 0.1838 | 0.0025 | -0.0014 |
| Sichuan | Chengdu | 29608 | 43.53 | -0.7592 | -0.8332 | 0.0027 | 0.0037 |

| Guizhou | Guiyang | 19710 | 36.41 | -1.2648 | -1.3716 | 0.0040 | 0.0066 |
| Yunnan | Kunming | 22195 | 39.31 | -1.1378 | -1.1523 | 0.0043 | 0.0047 |
| Shaanxi | Xian | 38564 | 50.02 | -0.3017 | -0.3424 | 0.0011 | 0.0010 |
| Gansu | Lanzhou | 21978 | 38.75 | -1.1489 | -1.1946 | 0.0058 | 0.0057 |
| Qinghai | Xining | 33181 | 47.44 | -0.5767 | -0.5376 | 0.0054 | 0.0046 |
| Ningxia | Yinchuan | 36394 | 50.67 | -0.4126 | -0.2933 | 0.0014 | 0.0005 |
| Xinjiang | Urumchi | 33796 | 43.98 | -0.5453 | -0.7992 | 0.0004 | 0.0006 |

**Note**: The unit of the level of urbanization is %, and the unit of GRP is *yuan* of Renminbi (RMB).

According to the theoretical model (*Results*), the analytical process of spatial crosscorrelation comprises three principal steps.

**Step1**: global analysis of spatial crosscorrelation. The basic measurement is the GSCI, which can be given by equations (9) and (10).

**Step2**: local analysis of spatial crosscorrelation. The basic measurements are the LSCIs, which can be calculated one by one using equations (17) and (18), or processed as batches using equations (19) and (20). In fact, two vectors of LSCIs can be visually displayed with a scatterplot.

**Step3**: explanation of spatial crosscorrelation scatterplots. Two pairs of scatterplots can be drawn using equations (27) to (32). Among them we need only one pair of scatterplots. Table 1 has shown the corresponding relationships between different equations.

## 4.2 Calculations and analyses

The new calculation method for Moran's index presented by Chen (2013) can be adapted to the SCI. Based on the standardized vector $x$, $y$ and the unitized weights matrix $W$, the SCI can be computed easily using Excel or Matlab. The method comprises three steps as follows. **Step 1**: **standardize the variables**. In other words, convert the initial vectors $X$, $Y$ in equation (1) into the standardized vectors in equation (5). As indicated above, the PSD instead of the SSD will be employed to standardize the data so that the results are comparable with Moran's index. The results of 2012 are shown in Table 3. **Step 2**: **unitize the spatial weight matrix**. Using the matrix of railway distances, we can compute the spatial contiguity matrix with the distance decay function $v(x)=1/x$, where $x$ denotes the railway distance between any two capital cities. Note that the diagonal elements of the matrix should be turned into zeros. Then unitize the contiguity matrix by using the sum of the whole entries to divide each entry. The final weights matrix can be expressed with equations (7) and (8). **Step 3**: **compute SCI**. According to equation (9), the USWM is first left multiplied by the transpose of $x$, and then the product of $x^T$ and $W$ is right multiplied by $y$; According to equation (10), the unitized weights matrix is first left multiplied by the transpose of $y$, and then the product of $y^T$ and $W$ is right multiplied by $x$. The final product of the continued multiplication yields the value of the SCI, and the two results are equivalent to one another. For example, in 2012, the index of spatial crosscorrelation between the level of urbanization and per capita GRP is $R_c=x^T W y \approx 0.1566$, $R_c=y^T W x \approx 0.1566$.

A pair of scatterplots of spatial crosscorrelation can be drawn using two approaches. The first approach is to make use of the variables $x$, $y$, $nWx$, $nWy$, $xx^T W y$, and $yy^T W x$. One scatterplot is based on the relationship between $x$ ($x$-axis) and $nWy$ as well as $xx^T W y$ ($y$-axis), which reflect the action of $x$ (per capita GRP) on $y$ (level of urbanization). The relationship between $x$ and $nWy$ gives the scatterpoints, while the relationship between $x$ and $xx^T W y$ yields the trendline (Figure 2a). The other scatterplot is based on the relationship between $y$ (horizontal axis) and $nWx$ as well as $yy^T W x$ (vertical axis), which reflect the reaction of $y$ (level of urbanization) on $x$ (per capita GRP). The relationship between $y$ and $nWx$ yields the scatterpoints, while the relationship between $x$ and $xx^T W y$ gives the trendline (Figure 2b).

The second approach is to utilize the variables $x$, $y$, $nWx$, $nWy$, $xy^T W x$, and $yx^T W y$. Compared with the first approach, $xx^T W y$ is replaced by $xy^T W x$, and $yy^T W x$ is substituted by $yx^T W y$. The results and effects are same as those from the first approach, and the scatterplots are the same as those displayed in Figure 2. In the scatterplots, the slopes of the trend lines equal the SCI value. This suggests that we can employ regression analysis based on the least squares method to estimate the SCI by using equations (21) and (22). If the independent variable is $x$, the dependent variable will be $nWy$. For 2012, the SCI value is about $R_c=0.1566$, and the coefficient of determination is approximately $R^2=0.2611$. This suggests that the per capita GRP can explain about 26.11% of the spatial change of the level of urbanization. If the independent variable is $y$, the dependent variable will be $nWx$. For 2012, the SCI value is still about $R_c=0.1566$, but the determination coefficient is approximately $R^2=0.1773$. This suggests that the level of urbanization can explain about 17.73% of the spatial change of the per capita GRP. Note that the intercept should be set to 0 as there is no constant term in the regression equations abovementioned. A discovery is that equation (21) and equation (22) give the same SCI value, but the values of goodness of fit are different. For our example, the action of $x$ on $y$ ($R^2$=0.2611) is stronger than the reaction of $y$ on $x$ ($R^2$=0.1773). This seems to suggest that the influence of economic development on urbanization is greater than the impact of urbanization on economic development.

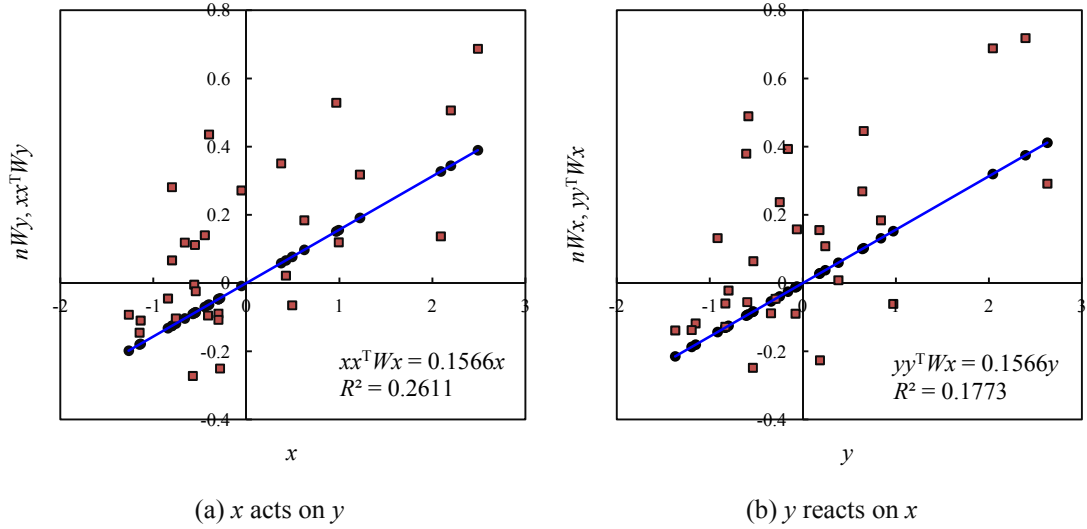

(a) $x$ acts on $y$          (b) $y$ reacts on $x$

**Figure 2 The scatterplots of spatial crosscorrelation between the per capita GDP and the level of urbanization in cities of China (2012)**

The coefficient of simple correlation between the level of urbanization and that of economic development of Mainland China can be decomposed by using the SCI value. For 2012, the simple correlation coefficient can be calculated with equation (35), and the result is about $R_0$=0.9457. In fact, it is easy to obtain the $R_0$ value in MS Excel using the function "correl" or "pearson", and the expression is "=correl($x$, $y$)" or "=pearson($x$, $y$)". Thus, according to equation (36), the PSCC is approximately $R_p$=0.9457-0.1566=0.7891. A conclusion can be drawn from these values of correlation coefficients that the direct correlation index of the 29 regions is about 0.7891, and the indirect correlation index is about 0.1566. The former has no relation to distances between different cities and can be regarded as intragroup correlation, and the latter is related to spatial interaction of different regions based on distances and can be treated as intergroup correlation.

Further, the SCI can be separated into LSCI, which reflect the spatial correlation between a region or city and all other regions or cities. Using equation (19), we can calculate the first vector of the local spatial correlation coefficient, which reflects the action $x$ (economic development) on $y$ (urbanization); using equation (20), we can compute the second vector of LSCI, which reflects the reaction *y* (urbanization) on *x* (economic development). All the results are displayed in Table 3, which shows that the sum of the LSCI equals the GSCI.

A scatterplot of local spatial crosscorrelation can be drawn by using the two sets of LSCI values (Figure 3). The scatterplot can be used to categorize Chinese regions in terms of spatial crosscorrelation. All the 29 regions can be classified into 4 groups according to the quadrants of a Cartesian coordinate system. For example, Beijing, Tianjin, and Shanghai are in the first quadrant, Heilongjiang is in the second quadrant, Guangdong, Hebei, Shanxi are in the third quadrant, and Chongqing and Shandong are in the fourth quadrant. For Beijing and Shanghai, the action of economic development on urbanization is weaker than the reaction of urbanization on economic development, but for Tianjin, Jiangsu, Zhejiang, and Fujian, the action of economic development on urbanization is stronger than the reaction of urbanization on economic development.

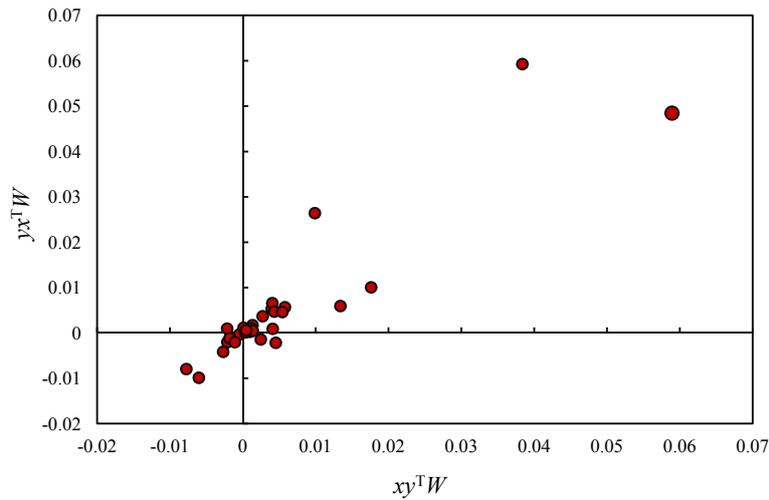

**Figure 3 The scatterplot of local spatial crosscorrelation between the per capita GRP and the level of urbanization in regions of China (2012)**

The analytical process of spatial crosscorrelation can be applied to the datasets of the years from 2000 to 2012. The calculation results include SCC ($R_0$), SCI ($R_c$), and PSCC ($R_p$), and the goodness of fit for the regression analyses of spatial crosscorrelation have been estimated (Table 4). From these calculations, we can get useful spatio-temporal information for China's urbanization and economic development. **First, the spatial crosscorrelation between per capita GRP and the level of urbanization became stronger and stronger.** The simple correlation is relatively stable, and the SCC values fluctuate around 0.95. However, the SCI values went up and up, while the PSCC values went down gradually (Figure 4). This suggests that the spatial interaction between different regions and cities became more and more significant in the process of spatio-temporal evolution of regional systems. **Second, the action of economic development on urbanization is stronger than the reaction of urbanization on economic development.** The goodness of fit for the regression of *nWy* depending on *x*, $R^2_{(y-x)}$, is all greater than that for the regression of *nWx* depending on *y*, $R^2_{(x-y)}$. This suggests that economic development is a cause of urbanization, and urbanization is an effect of economic development. Both the values of $R^2_{(x-y)}$ and $R^2_{(y-x)}$ go up and up from 2000 to 2012. This lends further support to the inference that the spatial interaction of the 29 regions became more and more significant over time. The absolute growth rate of $R^2_{(y-x)}$ is less than that of $R^2_{(x-y)}$. However, the relative growth rate of $R^2_{(y-x)}$ is greater than that of $R^2_{(x-y)}$, which can be shown by the allometric relationship between $R^2_{(x-y)}$ and $R^2_{(y-x)}$ (Figure 5). The allometric scaling exponent of $R^2_{(y-x)}$ depending on $R^2_{(x-y)}$ is about 1.655, which is significantly greater than 1. This lends further support that the level of urbanization in a geographical region is determined by the level of economic development and in turn reacts to it.

**Table 4 The values of SCC, SCI, PSCC and determination coefficients of the 29 Chinese regions (2000-2012)**

| Year | 2000 | 2005 | 2006 | 2007 | 2008 | 2009 | 2010 | 2011 | 2012 |
|---|---|---|---|---|---|---|---|---|---|
| SCC $R_0$ | 0.9142 | 0.9451 | 0.9447 | 0.9470 | 0.9523 | 0.9512 | 0.9577 | 0.9520 | 0.9457 |
| SCI $R_c$ | 0.0995 | 0.1382 | 0.1409 | 0.1415 | 0.1521 | 0.1550 | 0.1566 | 0.1575 | 0.1566 |
| PSCC $R_p$ | 0.8147 | 0.8068 | 0.8038 | 0.8056 | 0.8001 | 0.7962 | 0.8011 | 0.7945 | 0.7891 |
| $R^2_{(x-y)}$ | 0.0929 | 0.1991 | 0.2064 | 0.2011 | 0.2460 | 0.2469 | 0.2755 | 0.2716 | 0.2611 |
| $R^2_{(y-x)}$ | 0.0288 | 0.0958 | 0.1059 | 0.1106 | 0.1355 | 0.1484 | 0.1601 | 0.1734 | 0.1773 |
| $R^2_{(x-y)}+R^2_{(y-x)}$ | 0.1217 | 0.2949 | 0.3123 | 0.3117 | 0.3815 | 0.3953 | 0.4356 | 0.4450 | 0.4384 |

**Note**: The statistical data of the level of urbanization from 2001 to 2004 are absent in the website of China's NBS. The statistic $R^2_{(x-y)}$ denotes the goodness of fit for the regression of *nWy* depending on *x*, and $R^2_{(y-x)}$ refers to the goodness of fit for the regression of *nWx* depending on *y*.

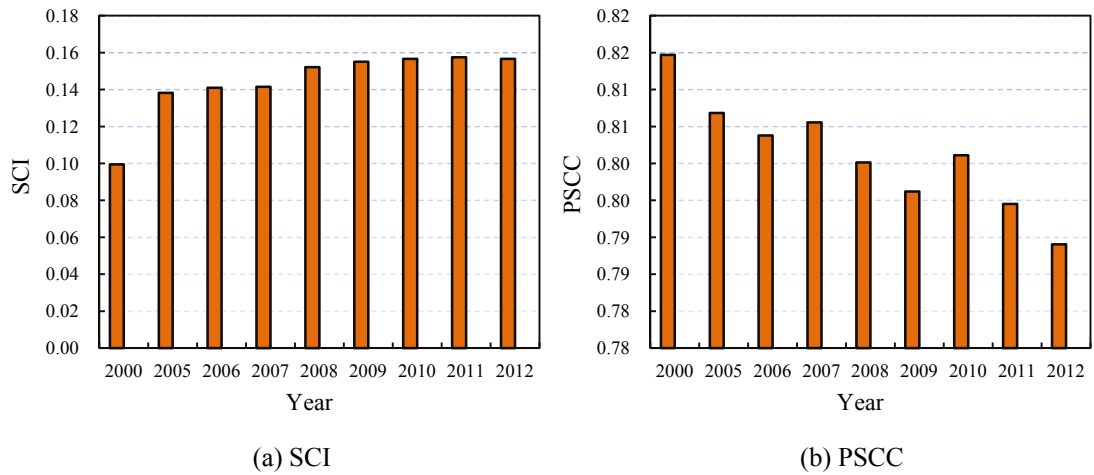

(a) SCI  (b) PSCC

**Figure 4 Histograms of SCI and PSCC of the spatial crosscorrelation of 29 Chinese regions**

**(2000-2012)**

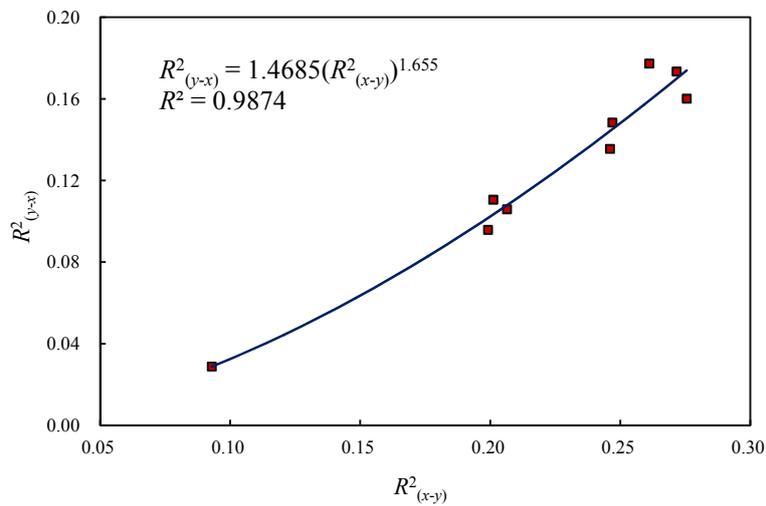

$$R^2_{(y\text{-}x)} = 1.4685(R^2_{(x\text{-}y)})^{1.655}$$
$$R^2 = 0.9874$$

**Figure 5 The allometric relationship between two kinds of determination coefficients for spatial**

**crosscorrelation analysis (2000-2012)**

# 5 Conclusions

The theory of spatial correlation analysis needs both spatial autocorrelation and spatial crosscorrelation measurements; the latter provides a new framework for geographical research. Based on the theoretical results and empirical studies, three basic conclusions can be drawn as follows.

**First, spatial crosscorrelation and spatial autocorrelation can complement one another.** Both spatial autocorrelation and spatial crosscorrelation analyses can be employed to study different geographical elements in a regional system or different subsystems within a geographical system. The two methods are different, but they can combine to make an integrated framework. The spatial autocorrelation analysis shows the simultaneous change in value of one numerically valued random variable, while the spatial crosscorrelation analysis displays the simultaneous change in values of two random variables. If we use one variable to measure a number of spatial entities, we can utilize spatial autocorrelation analysis; on the other hand, if we use two or more variables to measure a number of spatial entities, we can utilize both spatial autocorrelation analysis and spatial crosscorrelation analysis.

**Second, the spatial crosscorrelation coefficient represents the indirect relationships between spatial variables.** Using SCI, we can analyze the well-known simple correlation coefficient in spatial statistics. Pearson's correlation between two spatial variables includes two components: direct correlation and indirect correlation. The spatial correlation coefficient reflects the indirect correlation based on the spatial contiguity between any two geographical entities. Pearson's correlation coefficient minus the spatial crosscorrelation coefficient leaves the direct correlation coefficient. The direct correlation is actually a kind of partial correlation, which is independent of spatial patterns. In this sense, spatial crosscorrelation analysis can reveal the importance of the part played by geographical distances or spatial relationships.

**Third, the scatterplots of spatial crosscorrelation can be used to reveal the causality between two variables.** Pearson's correlation coefficient and spatial crosscorrelation coefficient can reflect the correlation between two variables, but they cannot distinguish between cause and effect. Fortunately, the scatterplots of spatial crosscorrelation can be used to differentiate between the cause and the effect. The spatial crosscorrelation plots appear by twos, and the two plots are of asymmetry. Therefore, they can show us which variable is in the leading position and which is in the subordinate position. In scientific research, determining causality may be more important than describing correlation in a geographical system. In this sense, the new framework of spatial crosscorrelation can be useful in future spatial analyses.

**Acknowledgement:**

This research was sponsored by the National Natural Science Foundation of China (Grant No. 41171129. See: https://isis.nsfc.gov.cn/portal/index.asp).

Legendre (Eds). *Developments in Numerical Ecology, NATO ASI Series, Vol. G14*. Berlin: Springer-Verlag, pp431-466

Stark JH, Sharma R, Ostroff S, Cummings DAT, Ermentrout B, *et al*. (2012). Local spatial and temporal processes of influenza in Pennsylvania, USA: 2003–2009. *PLoS ONE*, 7(3): e34245

Tiefelsdorf M (2002). The saddle point approximation of Moran's $I$ and local Moran's $I_i$ reference distributions and their numerical evaluation. *Geographical Analysis*, 34(3): 187-206

Wall MM (2004). A close look at the spatial structure implied by the CAR and SAR models. *Journal of Statistical Planning and Inference*, 121(2): 311-324

Wang J (2006). *Spatial Analysis*. Beijing: Scientific Press [In Chinese]

Wang JF, Reis BY, Hu MG, Christakos G, Yang WZ, *et al* (2011). Area disease estimation based on sentinel hospital records. *PLoS ONE*, 6(8): e23428

Wang JF, Stein A, Gao BB, Ge Y (2012). A review of spatial sampling. *Spatial Statistics*, 2(1): 1-14

Weeks JR, Getis A, Hill AG, Gadalla MS, Rashed T (2004). The fertility transition in Egypt: Intraurban patterns in Cairo. *Annals of the Association of American Geographers*, 94(1): 74-93

Zhou YX (1989). On the relationship between urbanization and gross national product. *Chinese Sociology & Anthropology*, 21(2): 3-16
24